\documentclass{jps-cp}
\usepackage{txfonts} 
\usepackage{bm}

\title{Dynamical Mean-Field Theory of Strongly Correlated Electron Systems}

\author{Dieter Vollhardt}

\inst{Theoretical Physics III, Center for Electronic Correlations and Magnetism, Institute of Physics, University of
Augsburg, D-86135 Augsburg, Germany}

\email{Dieter.Vollhardt@physik.uni-augsburg.de}

\recdate{September 4, 2019}

\abst{Dynamical Mean-Field Theory (DMFT) has opened new perspectives for the investigation of strongly correlated electron systems and greatly improved our understanding of correlation effects in models and materials.
In contrast to Hartree-Fock-type approximations the
mean field of DMFT is dynamical, whereby local quantum
fluctuations are fully taken into account.
DMFT becomes exact in the limit of high spatial dimensions or coordination number.
Using DMFT the dynamics of correlated electron systems can be investigated
non-perturbatively at all interaction strengths, electron densities and temperatures.
By merging density functional theory with DMFT a powerful method for the calculation
of the properties of correlated electron materials has become available, which is applicable
to bulk systems and heterostructures, including topological states of matter.
 The inclusion of non-local correlations into DMFT makes it possible to explore unconventional superconductivity and the critical behavior at thermal or quantum phase transitions. By generalizing DMFT to non-equilibrium states also
the real-time dynamics of correlated systems can be investigated.
In this brief review the foundations and current state of DMFT are discussed along with characteristic physical insights obtained with this approach.}

\kword{Dynamical mean-field theory, correlated electron systems, correlated materials}

\begin{document}
\maketitle

\section{Electronic correlations}

Nature is characterized by correlations in space and time.
In particular, in solids correlations between electrons play an important role. They denote electronic interaction effects which cannot be explained within a single-electron picture as obtained by factorization approximations such as Hartree-Fock theory.
Electronic correlations are known to lead to the emergence of complex phenomena
which include the Mott metal-insulator transition \cite{Mott+Tokura}, heavy fermion behavior \cite{Heavy-fermions}, high-temperature superconductivity \cite{Schrieffer}, colossal magnetoresistance \cite{Dagotto}, and Fermi liquid instabilities \cite{HvL}.
Such properties are not only of interest for fundamental research, but also for technological applications.
Indeed, the exceptional sensitivity of correlated electron materials with respect to
changes of external parameters
(temperature, pressure, magnetic
and/or electric fields, doping, etc.)
can be employed to develop materials with useful functionalities \cite{Functionality}.
Consequently there is a great need for investigation techniques which enable accurate theoretical explorations of correlated electron systems \cite{Juelich-lecture-notes1}.

The single-band Hubbard model \cite{Gutzwiller-HubbardI-Kanamori} is the prototypical, and simplest, microscopic lattice model of interacting electrons in a solid such as $3d$ electrons in transition metals.
%
The Hamiltonian  consists of a kinetic energy
and an interaction which is assumed to be purely local, i.e., extremely screened (due to the Pauli principle the two interacting electrons must have opposite spin):
\begin{equation}
\hat{H} =  \sum_{\bm{R}_i , \bm{R}_j} \sum_{\sigma}
t_{ij}  \hat{c}_{i \sigma}^{+} \hat{c}_{j \sigma}^{}  + U \sum_{\bm{R}_i} \hat{n}_{i \uparrow} \hat{n}_{i \downarrow}.
\label{G11.7c}
\end{equation}
Here $t_{ij}$ is the amplitude for  hopping between sites $\bm{R}_i$ and $\bm{R}_j$, $U$ is the local Coulomb interaction, $\hat{c}_{i \sigma}^{+} (\hat{c}_{i \sigma}^{})$ are creation (annihilation)
operators of electrons with spin $\sigma$ at site $\bm{R}_i$, and
$\hat{n}_{i \sigma}^{} = \hat{c}_{i \sigma}^{+} \hat{c}_{i \sigma}^{}$  (operators are denoted by a hat). The Fourier transform
of the kinetic energy
$\hat{H}_{kin} = \sum_{\bm{k} , \sigma}
\epsilon_{\bm{k}} \hat{n}_{\bm{k} \sigma}^{}$
involves the dispersion
$\epsilon_{\bm{k}}$ and the momentum distribution
$\hat{n}_{\bm{k} \sigma}^{}$.
%
In the atomic limit ($t_{ij}=0$) a lattice site can either be empty, singly occupied (spin up or down), or doubly occupied (spin up and down). These three states are then local eigenstates. A finite kinetic energy ($t_{ij} \neq 0$) leads to transitions between the three states on each lattice site as indicated in fig.~\ref{DMFT}.
 These local quantum fluctuations are independent of the dimensionality and lattice structure of the system and have no classical counterpart.

Analytic solutions of the Hubbard model are not available in those dimensions which are of particular interest in solid-state physics ($d = 2,3$), while numerical solutions are limited
by the exponential increase of the Hilbert space dimension with the number of particles.
This calls for comprehensive non-perturbative approximation schemes which are applicable for all input parameters.

\subsection{Mean-field theories}
\label{MFT}

For classical and quantum-mechanical many-particle models an approximate,
overall description of their properties can often be obtained
within a mean-field theory (MFT). While in the actual model each particle or spin experiences a complicated, fluctuating field generated by the interaction with other particles or spins, in a MFT this field is approximated by an average (``mean'') field.

A MFT can be constructed, for example, by factorizing the interaction. Alternatively one can make some variable or parameter large (infinite), whereby fluctuations are  suppressed. Depending on the model this can be the spin $S$, the interaction range, the spin degeneracy $N$, the spatial dimension $d$, or the number of nearest neighbors of a lattice site (coordination number) $Z$.
The best-known MFT in statistical physics is the Weiss
molecular-field theory for the Ising model.
It
becomes exact in infinite dimensions $d$ or coordination
number $Z$ (with $Z=2d$ for hypercubic lattices). For the energy to remain finite in this limit
the nearest-neighbor coupling $J$ needs to be rescaled as $J \to  J^*/Z$, with $J^* = {\rm const}$
(``classical scaling'').
By contrast, lattice fermion models such as the Hubbard model are more intricate than localized spin models.
Therefore the construction of a MFT  with the
comprehensive properties of the Weiss MFT for the Ising
model is also more complicated.
The simplest MFT of the Hubbard
model is the Hartree approximation,
which factorizes the local interaction  $\hat{n}_{i \uparrow} \hat{n}_{i \downarrow}$ (since the two spins are opposite there is no Fock term). Thereby correlations are eliminated and the mean field becomes static. Hence a factorization approximation cannot describe correlation effects in the Hubbard model.

\section{Dynamical mean-field theory (DMFT)}

The limit of high spatial dimensions $d$ or coordination number $Z$ can also be used to construct a generic MFT for Hubbard-type models.
The construction is made possible by the fact that  diagrammatic quantum many-body perturbation theory greatly simplifies for $d$ or $Z \to \infty$, since one-particle irreducible diagrams in position space collapse in this limit, implying that only local diagrams remain \cite{MV89}.
The diagrammatic collapse can be understood as follows \cite{metzner89a}: the probability for a quantum particle to hop from a site $\bm{R}_i$ to some nearest-neighbor site $\bm{R}_j$ is ${\cal O}(1/Z)$. The corresponding amplitude, and therefore the one-particle propagator $G_{ij}$,  is then ${\cal O} (1/\sqrt{Z})$, such that the local propagator $G_{ii}$ is the largest term.
For the energy to remain finite in this limit
the nearest-neighbor hopping amplitude $t$ must be rescaled as $t \to  t^*/\sqrt{Z}$, with $t^* = {\rm const}$  \cite{MV89}
(``quantum scaling'').

These diagrammatic simplifications make it possible to derive
a comprehensive, diagrammatically controlled MFT for lattice fermions \cite{Janis91,Georges92,Jarrell92}; for detailed discussions see refs. \cite{georges96,dmft_phys_today,Vollhardt-Salerno}.
The self-energy is momentum independent, i.e., local, but retains its frequency dependence \cite{MH89a}. Thus it continues to describe the many-body dynamics of the interacting system.
Consequently, the local propagator $G_{ii}(\omega)$ and the local self-energy $\Sigma_{ii}(\omega)$ are the determining quantities in this limit. As in the case of the Weiss MFT for the Ising model, the actual lattice fermion problem therefore reduces to an effective single-site problem where, however, the mean field is dynamical.
This is in contrast to Hartree-Fock theory where the self-energy acts only as an additional static potential.
The resulting problem is mean-field-like  \emph{and} dynamical and is therefore a \emph{dynamical mean-field theory}  (DMFT) for lattice fermions which is able to describe genuine correlation effects as will be discussed next.

\subsection{The self-consistent DMFT equations}
\label{DMFT-equations}

\begin{figure}
\centerline{\includegraphics[width=0.8\textwidth]{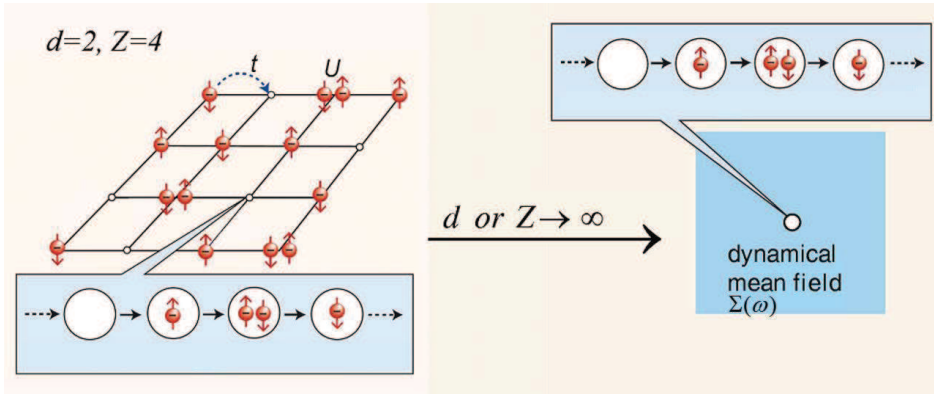}}
\caption{Left: Hubbard model on a square lattice ($d=2$, $Z=4$); the characteristic local fluctuations of the electrons are indicated schematically. Right: In the limit $d$ or $Z\rightarrow \infty $ the Hubbard model \index{Hubbard model} effectively reduces to a dynamical single-site problem which may be viewed as a lattice site embedded in a $\bm{k}$-independent, but dynamical, fermionic mean field.  Electrons can hop from the mean field onto this site and back, and interact on the site as in the original Hubbard model. The local dynamics of the electrons is independent of the dimension or coordination number and therefore remains unchanged.}
\label{DMFT}
\end{figure}

DMFT can be derived in different ways, not only mathematically, but also regarding the physical interpretation of the correlation problem emerging for $d,Z \to \infty$ \cite{Janis91,Georges92}.
The derivations make use of the fact that in $d=\infty$  lattice fermion models with a local interaction effectively reduce to a
single site embedded in a dynamical mean field provided by the other fermions \cite{georges96,dmft_phys_today,Vollhardt-Salerno} as illustrated in fig.~\ref{DMFT}.
The derivation of DMFT by a self-consistent mapping of
the $d$-dimensional lattice problem  onto a single-impurity Anderson model
 \cite{Georges92} (see also \cite{Jarrell92})
proved to be a physically intuitive and particularly practical approach since it connects with the well-studied theory of quantum impurities for whose investigation efficient numerical codes such as the quantum Monte-Carlo (QMC) method \cite{Hirsch86} already existed.

At finite temperatures the DMFT equations consist of the expression for the local propagator $G_{\sigma }(i\omega _{n})$, where $\omega_n = (2n+1) \pi T$ are Matsubara frequencies and we dropped the local site indices $ii$ and added the spin index $\sigma$, and a self-consistency condition. The local propagator can be written as
\begin{equation}
G_{\sigma }(i\omega _{n})=-\frac{1}{{\mathcal{Z}}}\int \prod_{\sigma
}Dc_{\sigma }^{\ast }Dc_{\sigma }[c_{\sigma }(i\omega _{n})c_{\sigma }^{\ast
}(i\omega _{n})]\exp [-S_{\mathrm{loc}}],
\label{Vollhardt:impurity_problem}
\end{equation}
with the partition function ${\mathcal{Z}}=\int \prod_{\sigma }Dc_{\sigma }^{\ast }Dc_{\sigma }\exp [-S_{\mathrm{loc}}]$
and the local action  \\
$S_{\mathrm{loc}}=
- \int_0^{\beta} d \tau_1 \int_0^{\beta} d \tau_2
\sum_{\sigma} c^{*}_{\sigma} (\tau_1) \mathcal{G}^{-1}_{\sigma}(\tau_1-\tau_2) c _{\sigma}(\tau_2)
+ \; U\int_0^{\beta} d \tau
c^{*}_{\uparrow}(\tau)c_{\uparrow}(\tau)c^{*}_{\downarrow}(\tau)c_{\downarrow}(\tau)$.
Here $\mathcal{G}_{\sigma}$ is the effective local propagator (also called ``bath Green function'', or ``Weiss mean field'' \cite{georges96}),
which is defined by the Dyson equation
$\mathcal{G}_{\sigma}(i\omega _{n}) = \lbrack \lbrack G_{\sigma }(i\omega _{n})]^{-1} + \Sigma _{\sigma }(i\omega _{n})]^{-1}$.
In principle, both $\mathcal{G}_{\sigma
}(i\omega _{n})$ and $\Sigma _{\sigma }(i\omega _{n})$
can be viewed as a local, dynamical
mean field since both
appear in the bilinear term of the local action.
By identifying the
local propagator (\ref{Vollhardt:impurity_problem}) with the local lattice Green function (the Hilbert transform of the lattice Green function
$G_{\bm{k}\,\sigma }(i\omega _{n})=\lbrack i\omega _{n}-\epsilon _{\bm{k}}+\mu -\Sigma _{\sigma }(i\omega _{n})]^{-1}$),
which is an exact property in $d=\infty $ \cite{georges96}, one obtains the self-consistency condition
\begin{equation}
G_{\sigma }(i\omega _{n}) = \frac{1}{L} \sum_{\bm{k}}G_{\bm{k}\,\sigma }(i\omega_{n}) =
\underset{-\infty }{\overset{\infty }{\int }}d\epsilon \frac{N(\epsilon)}{i\omega _{n}-\epsilon +\mu -\Sigma _{\sigma }(i\omega _{n})}.
\label{Vollhardt:local_prop-1}\\
\end{equation}
Although DMFT corresponds to an effectively local problem, the propagator $G_{\bm{k}}(\omega)$ depends on the crystal momentum $\bm{k}$, but only through the dispersion relation $\epsilon_{\bm{k}}$ of the non-interacting electrons. There is no additional momentum-dependence through the self-energy, since this quantity is local within DMFT.
In (\ref{Vollhardt:local_prop-1}) the ionic lattice enters only through the density of states (DOS) $N(\epsilon)$ of the non-interacting electrons. This equation implies
$G_{\sigma }(i\omega _{n})= G_{\sigma }^{0}(i\omega _{n}- \Sigma _{\sigma }(i\omega _{n}))$, which
illustrates the mean-field character of the DMFT particularly clearly: the local Green function  of the interacting system corresponds to the non-interacting Green function $G_{\sigma }^{0}$ at the renormalized energy $i\omega _{n}- \Sigma _{\sigma }(i\omega _{n})$, which is the energy measured relative to the interaction-induced mean-field energy $\Sigma _{\sigma }(i\omega _{n})$ of the surrounding dynamical fermionic bath.

In the self-consistent DMFT equations each frequency is coupled to all other frequencies. This shows that DMFT is still a full-scale many-body theory. The solution of the self-consistent equations requires the application of powerful numerical methods, in particular quantum Monte-Carlo (QMC) simulations \cite{Jarrell92,georges96,QMC}, with continuous-time QMC \cite{CT-QMC2011} as the method of choice, the numerical renormalization group \cite{NRG-RMP}, the density matrix renormalization group \cite{DMRG}, exact diagonalization \cite{georges96} and Lanczos procedures \cite{ED+Lanczos}.

\subsection{Characteristic features of DMFT}

In DMFT the mean field is dynamical, whereby local quantum
fluctuations are fully taken into account, but is local (i.e., spatially independent) because of the infinitely many neighbors of every lattice site (``single-site DMFT'').
The only approximation of DMFT when applied in $d<\infty$ is the neglect of the $\bm{k}$-dependence of the self-energy. DMFT
provides a comprehensive, non-perturbative, thermodynamically consistent and diagrammatically controlled approximation scheme for the investigation of correlated lattice models at all interaction strengths, densities, and temperatures \cite{georges96,dmft_phys_today}, which
 can resolve even low energy scales.
It describes fluctuating moments, the renormalization of quasiparticles and their damping, and is especially valuable for the study  of correlation problems at intermediate couplings, where no other methods are available.
Unless a symmetry is broken the $\bm{k}$-independence of the self-energy implies typical Fermi-liquid properties of the DMFT solution \cite{MH89b}.

Most importantly, DMFT makes it possible to compute electronic correlation effects quantitatively in such a way that they can be tested experimentally, for example, by electron spectroscopies (see below). Namely, DMFT describes the correlation--induced transfer of spectral weight and the finite lifetime of quasiparticles through the real and imaginary part of the self-energy, respectively.
This greatly helps to understand and characterize the Mott metal-insulator transition (MIT).

\subsection{Applications of DMFT}
\subsubsection{Mott-Hubbard metal-insulator transition}
\label{sec:mit}

The interaction-driven transition between a paramagnetic metal and a
paramagnetic insulator, first discussed by Mott \cite{mott} and referred to as Mott-MIT (or Mott-Hubbard MIT when studied within the Hubbard model), is  one of the most intriguing phenomena in
condensed matter physics~\cite{Mott+Tokura}.
This transition is
a consequence of the quantum-mechanical competition
between the kinetic energy of the electrons and their local interaction $U$: the kinetic energy prefers the electrons to be mobile (a wave effect) which invariably leads to their interaction (a particle effect). For large values of $U$ doubly occupied sites become energetically too costly. The system then reduces its total energy by localizing the electrons, which leads to a MIT.
Here the DMFT has been extremely valuable since it provides detailed insights into the nature of the Mott-Hubbard-MIT  for all values of the interaction $U$ and temperature $T$ \cite{georges96,dmft_phys_today,Vollhardt-Salerno}. A comprehensive microscopic analysis of the Mott MIT within Fermi liquid theory as derived by DMFT has been performed recently \cite{Krien2019}.

While at small $U$ the system can be described by coherent quasiparticles
whose DOS still resembles that of the free electrons, the spectrum in the Mott
insulating state consists of two separate incoherent ``Hubbard
bands'' whose centers are separated approximately by the energy
$U$ (here we discuss only the half filled case without magnetic order).
At intermediate values of $U$ the spectrum then has a
characteristic three-peak structure which is qualitatively similar to that of the single-impurity
Anderson model \cite{SIAM} and which
is a consequence of
the three possible occupations of a lattice site: empty, singly occupied (up or down), doubly occupied.
At $T=0$ the width of the quasiparticle peak vanishes at a critical value of $U$ which is of the order of the band width.
This shows that the Mott-Hubbard MIT is an intermediate-coupling problem and therefore belongs to the hard problems in many-body theory.
At $T>0$ the Mott-Hubbard MIT  is of first order
and is associated with a hysteresis region in
the interaction range $U_{\rm c1}<U<U_{\rm c2}$ where $U_{\rm c
1}$ and $U_{\rm c 2}$ are the values at which the insulating and
metallic solution, respectively, vanish
\cite{georges96,MIT}; for a detailed discussion see refs.\cite{georges96,dmft_phys_today,Vollhardt-Salerno}.
The hysteresis region
terminates at a critical point,
above which
the transition becomes a smooth crossover from
a ``bad metal'' to a ``bad insulator''; for a schematic plot of the phase diagram see fig. 3 of ref. \cite{dmft_phys_today}.
Transport in the incoherent region above the critical point shows remarkably rich properties, including scaling behavior
\cite{Transport-MIT}.

 Mott-Hubbard MITs are found, for example, in transition metal oxides  with
partially filled bands. For such systems  band theory
typically predicts metallic behavior. One of the most famous  examples is V$_{2}$O$_{3}$ doped with Ti or Cr~\cite{V2O3}.
However, it is now known that certain organic materials
are better realizations of the single-band Hubbard model
without magnetic order and
allow for much more controlled investigations of the Mott state and the Mott MIT \cite{MIT-in-organics}.

\subsubsection{Metallic ferromagnetism}
\label{Metallic-ferromagnetism}

The Hubbard model had been introduced in 1963 \cite{Gutzwiller-HubbardI-Kanamori} as an attempt to explain metallic ferromagnetism in 3$d$ metals such as Fe, Co, and Ni starting from a minimal microscopic model. However, at that time the problem could not be solved. Therefore  it was uncertain for a long time whether the Hubbard model can explain band ferromagnetism at realistic temperatures, electron densities, and interaction strengths in dimensions $d>$1 at all.
Three decades later it was shown within DMFT  that the Hubbard model on a generalized fcc lattice indeed describes metallic ferromagnetic phases in large regions of the phase diagram \cite{ferromagnetism}.
In the paramagnetic phase the
susceptibility $\chi _{F}$ obeys a Curie-Weiss law, where the Curie temperature $T_{C}$  is now much lower than that obtained within Stoner theory due to many-body effects. In the ferromagnetic phase the magnetization $M$
is consistent with a Brillouin function as originally derived for localized spins, but for a \emph{non-integer}
magneton number  as in 3$d$
transition metals.
Therefore, DMFT accounts for the behavior of both the magnetization and the susceptibility in
band ferromagnets. Metallic ferromagnetism is seen to be another intermediate coupling problem.

\subsubsection{Disorder}
\label{Electronic Correlations and Disorder}

DMFT also provides a non-perturbative theoretical framework for the investigation of correlated electrons in the presence of disorder.
When the effect of local disorder is taken into account
through the arithmetic mean of the local DOS (LDOS) one obtains, in the
absence of interactions, the coherent potential approximation (CPA)
\cite{vlaming92+Janis92a}; for a discussion see ref. \cite{Vollhardt-Salerno}. However, CPA cannot describe Anderson
localization. To overcome this deficiency a variant of the DMFT was formulated where the
geometrically averaged LDOS is computed from the solutions of the
self--consistent stochastic DMFT equations and is then fed into the self--consistency cycle \cite{Dobrosavljevic97+03+15}.
Thereby a MFT of Anderson localization (``typical medium theory'') is obtained which
reproduces many of the known features of the disorder--driven
MIT for non--interacting electrons.
This scheme can be integrated into DMFT to study the properties of disordered electrons in the presence of
interactions and to compute, for example, the phase diagram of the
Anderson-Hubbard model   \cite{Byczuk05+Byczuk10}.

\section{DMFT for materials with correlated electrons}

\subsection{DFT+DMFT and $GW$+DMFT approach}
\label{LDA+DMFT}

The development and application of theoretical techniques to understand the basic features of the one-band Hubbard model took several decades.
During that time first-principles investigations of the much more complicated many-body problem posed by correlated materials were clearly out of reach.
The electronic properties of solids were  mainly studied within density-functional theory (DFT) \cite{DFT1-DFT2},
e.g., in the local density approximation (LDA) \cite{LDA}, the generalized gradient approximation (GGA) \cite{PB96}, and the LDA+U method \cite{Anisimov91}.
Those approaches can accurately describe the ground state properties of many simple elements and semiconductors, and even of some insulators. Moreover, they often correctly predict the magnetic and orbital properties \cite{LDA} as well as the crystal structures of many solids~\cite{Baroni01}.
However, these methods fail to describe the electronic
and structural properties of correlated paramagnetic materials since they miss characteristic features of correlated electron systems such as heavy quasiparticle behavior and Mott physics.
This changed dramatically with the advent of DMFT.
The computational scheme obtained by merging material-specific DFT-based approximations with the many-body DMFT \cite{Anisimov97+Lichtenstein98}, now referred to as DFT+DMFT (or more specifically LDA+DMFT, GGA+DMFT, etc.),
provides a powerful new method for the microscopic computation
of the electronic, magnetic, and structural properties of correlated materials
from first principles even at
finite temperatures \cite{psi-k+Held07,Kotliar06,Katsnelson08,FOR1346-Report-2017}. In particular, this approach naturally accounts for the existence of local moments in the paramagnetic phase.

As in the case of the Hubbard model the many-body model constructed within the DFT+DMFT scheme consists of two parts: an effective kinetic energy obtained by DFT which describes the material-specific band structure of the uncorrelated electrons, and the local interactions between the electrons in the same orbital as well as in different orbitals. Here the static contribution of the electronic interactions already included in the DFT-approximations must be subtracted to avoid double counting  \cite{Anisimov97+Lichtenstein98,psi-k+Held07,Kotliar06,Katsnelson08,FOR1346-Report-2017}. Such a correction is not necessary in the fully diagrammatic, but computationally very demanding $GW$+DMFT approach, where the LDA/GGA input is replaced by the $GW$ approximation \cite{GW+DMFT}. The complicated many-particle problem  obtained in this way with its numerous energy bands and local interactions is then solved within DMFT, typically by CT-QMC. By construction, DFT+DMFT includes the correct
quasiparticle physics and the corresponding energetics, and reproduces the DFT results in the limit of weak
Coulomb interaction $U$.
More importantly, DFT+DMFT describes the
correlation-induced dynamics of strongly correlated electron materials since it is able to account for the physics at all values of the
Coulomb interaction and doping.

The application of DFT+DMFT made investigations of correlated materials much more realistic. This enhanced realism led to the discovery of novel physical mechanisms and correlation phenomena. One example is the Mott MIT. Originally, using the single-band Hubbard model, the Mott MIT was explained as a transition where the effective mass of quasiparticles diverges (``Brinkman-Rice scenario'') \cite{Brinkman-Rice}. After DMFT had opened the way to study multi-band models, an ``orbital selective'' Mott MIT was identified \cite{Georges+Biermann}. Then, with the advent of DFT+DMFT, a ``site selective'' Mott MIT was found in Fe$_{2}$O$_{3}$ \cite{Leonov+Abrikosov}.
We now illustrate the DFT+DMFT approach by its application to two paradigmatic materials, SrVO$_3$ and Fe.

\subsection{SrVO$_3$: three-peak spectral function}

Transition metal oxides are an ideal laboratory for the study of
electronic correlations in solids.
Spectroscopic studies
typically find  a pronounced lower Hubbard band in the photoemission spectra
which cannot be explained by conventional band-structure theory.
SrVO$_{3}$ is a particularly simple correlated material due to its 3$d^{1}$ configuration and its purely cubic crystal structure with one vanadium ion per unit cell. 
The cubic symmetry of the crystal field splits the fivefold degenerate 3$d$ orbital into a threefold degenerate t$_{2g}$ orbital and an energetically higher twofold degenerate e$_{g}$ orbital. In the simplest approximation only the local interaction between the electrons in the t$_{2g}$ orbitals is included. By employing a variant of the LDA it is possible to compute the strength of the local Coulomb repulsion ($U\simeq $ 5.5 eV) and the Hund's rule coupling ($J\simeq $
1.0 eV) \cite{Sekiyama03+Nekrasov05+06}. Using these values the electronic band structure is then calculated within LDA. The correlated electron problem defined in this way is solved numerically within DMFT.
The spectral function obtained thereby shows the characteristic three-peak structure of a correlated metal (lower Hubbard band, quasiparticle peak, upper Hubbard band) \cite{Sekiyama03+Nekrasov05+06,Pavarini04}.
This was confirmed experimentally using electron spectroscopies \cite{Sekiyama03+Nekrasov05+06,Inoue94}.
Recently it was found that oxygen vacancy states created by UV or x-ray irradiation can strongly affect the line shapes of the lower Hubbard band and the quasiparticle peak \cite{Backes16}. This must be taken into account in quantitative interpretations of those peaks.

\subsection{Fe: electronic correlations and structural stability}

Iron (Fe) exhibits a rich phase  diagram.
Under ambient conditions Fe is ferromagnetic and has a bcc crystal
  structure ($\alpha$ phase). At the Curie temperature $T_C\sim$ 1043 K the $\alpha$
  phase becomes paramagnetic but retains its bcc structure. Only when the temperature is further increased to $T_{\text{struct}} \sim$
  1185 K does a structural phase transition to a fcc structure
  ($\gamma$ phase) take place. At $T \sim$ 1670 K a transition to a second bcc structure ($\delta$ phase) occurs.
DFT+DMFT calculations for ferromagnetic bcc Fe provide a semi-quantitative description of several physical properties of that phase (at least sufficiently far from the Curie point) and demonstrate that electronic correlations play an important role \cite{lichtenstein01+Fe-PES}. This approach also clarified the microscopic origin of the magnetic exchange interactions in the ferromagnetic phase \cite{exchange-Fe}.

In these investigations the lattice structure was assumed to be given and fixed. The question regarding the stability of the lattice structure in the presence of electronic correlations was studied, for example, in the case of plutonium \cite{Dai} and iron at ambient pressure \cite{Leonov11+12+14a+Kunes2017}.
While DFT band-structure methods provide qualitatively
  correct results for several electronic and structural properties
  of
  iron~\cite{Singh91+Okatov09+Kormann12}, their application to the
bcc-to-fcc phase transition predicts a simultaneous transition of the
  structural and the magnetic state.
In fact, in the absence of the magnetization
  standard band-structure methods find bcc iron to be
  unstable~\cite{Hsueh02}.
  These discrepancies can been resolved by using DFT+DMFT to compute total energies 
  \cite{Leonov11+12+14a+Kunes2017,Park14}.
Thereby one obtains values of the lattice constant, unit cell volume, and bulk modulus of paramagnetic $\alpha$ iron which are in good quantitative agreement with experiment \cite{Leonov11+12+14a+Kunes2017}.
In particular,
  DFT+DMFT calculations of the equilibrium crystal structure
and phase stability of iron
find that the bcc-to-fcc phase transition indeed takes place at a temperature well above the magnetic transition (at about 1.3 $T_C$) and  correctly determine the phonon dispersion and Debye temperature \cite{Leonov11+12+14a+Kunes2017}.
This approach is also of interest for geophysical studies, namely to explore iron and nickel at
Earth's core conditions \cite{Earths-core}.

\subsection{From bulk matter to heterostructures and topological properties}

DFT+DMFT has been remarkably successful in the investigation
of correlated materials, including transition metals and their oxides, manganites, fullerenes, Bechgaard salts, $f$-electron materials, magnetic superconductors,  and Heusler alloys  \cite{psi-k+Held07,Kotliar06,Katsnelson08,FOR1346-Report-2017}.  In particular, the study of Fe-based pnictides and chalcogenides led to the new insight that in  metallic multi-orbital materials the intra-atomic exchange  $J$  can lead to strong correlations \cite{Hunds-metal}. Clearly DMFT-based approaches will be very useful for the future design of correlated materials \cite{Adler-Kotliar-2018}, e.g., materials with a high thermopower for thermoelectric devices
which can convert waste heat into electric energy \cite{thermo}.

Furthermore, DMFT studies of inhomogeneous systems continue to improve our understanding of correlation effects at surfaces and interfaces, in thin films and in multi-layered nanostructures \cite{Heterostructures-etc} and thereby provide valuable insights into potential functionalities of such structures and their application in electronic devices.
DMFT has been extended to study correlations also in finite systems such as nanoscopic
 conductors and molecules \cite{molecules}. Thereby many-body effects were shown to be important even in biological matter, e.g., in the kernel of hemoglobin and molecules with important biological functions \cite{Weber}.

With DMFT it has been possible to  predict topological states of correlation models such as the Kitaev model in a uniform magnetic field and the interacting Haldane-Hubbard model \cite{Topological1}, and to explore the emergence of topological properties of materials such as cerium  \cite{Topological2}. Correlation-induced topological effects can also arise from non-Hermitian properties of the
single-particle spectrum in equilibrium systems \cite{Topological3}.

\section{Beyond single-site DMFT}

DMFT has been a breakthrough for the investigation and explanation of correlation effects in models and materials, such as quasiparticle renormalization and damping, spectral transfer, the Mott MIT, orbital, charge, and magnetic ordering and correlation-induced lattice instabilities. Although a dynamical, local self-energy is an approximation in  $d<\infty$, experiments with cold atoms in optical lattices demonstrated that single-site DMFT is remarkably accurate in $d=3$ \cite{Cold-atoms}. A dynamical, local self-energy was also shown to be well justified in iron pnictides and chalcogenides \cite{local1} as well as in Sr$_{2}$RuO$_{4}$ \cite{local2}. However,  single-site DMFT can certainly not describe the critical behavior at thermal or quantum phase transitions or unconventional superconductivity. Indeed,
when correlation effects occur on the scale of several lattice constants,  non-local extensions of DMFT are indispensible.

There are several different strategies to include non-local correlations into the DMFT \cite{nonlocal}; in the following we discuss four approaches.
(i) \emph{Extended DMFT}: This is an early strategy to include intersite quantum fluctuations into DMFT, where the interaction strength of a nearest-neighbor density-density interaction is scaled such that its fluctuation part contributes even in the large $d$ limit, i.e., beyond the Hartree level \cite{EDMFT}.
(ii) \emph{Cluster extensions}: Here the dynamical cluster approximation (DCA) \cite{DCA} and the cellular DMFT \cite{CDMFT} are widely used methods. They map a lattice model onto a cluster of sites (rather than onto a single site), which is then self-consistently embedded in a dynamical mean field. Thereby it has become possible to compute, e.g., typical features of unconventional superconductivity in the Hubbard model in $d=2$ such as the interplay of antiferromagnetism and $d-$wave pairing as well as pseudogap behavior  \cite{Maier-cluster}, and signatures of Anderson localization in disordered systems \cite{Cluster-Anderson-localization}.
(iii) \emph{Diagrammatic generalizations}:  By using diagrammatic extensions of the DMFT, corrections to the local self-energy in terms of Feynman diagrams can be calculated. Here the dynamical vertex approximation (D$\Gamma$A) \cite{DGA} and the dual fermion theory \cite{DF} are powerful approaches. They already provided important new insights into the mechanism of superconductivity arising from purely repulsive interactions, e.g., in the two-dimensional Kondo lattice model \cite{DMFT-superconductivity-Otsuki}  and the Hubbard model \cite{DMFT-superconductivity-Held}. In particular, in the repulsive Hubbard model a specific set of local particle-particle diagrams was identified which describe a strong screening of the bare interaction at low frequencies. Thereby antiferromagnetic spin fluctuations are suppressed, which in turn reduces the pairing interaction. Thus dynamical vertex corrections are found to reduce  $T_c$ strongly \cite{DMFT-superconductivity-Held}. With these approaches one can also determine critical behavior,  not only in the vicinity of thermal phase transitions ($T>0$) \cite{criticality1} but also near quantum phase transitions ($T=0$) \cite{criticality2}.
(iv) \emph{DMFT plus functional renormalization group (fRG):} In this approach the fRG flow \cite{fRG} does not start from the bare action of the system, but rather from the DMFT solution \cite{DMFT+fRG}. Local correlations are thus included already from the beginning, and nonlocal correlations are generated by the fRG flow, as was demonstrated for the two-dimensional Hubbard model in the  limit of strong interactions \cite{DMFT+fRG}.

\section{Nonequilibrium DMFT}

The study of correlated electrons out of equilibrium by employing a generalization of the DMFT to nonequilibrium situations has become yet another fascinating new research area \cite{RMP-Noneq-DMFT2014}. Nonequilibrium DMFT is able to explain, for example, the results of time-resolved  electron spectroscopy experiments, where femtosecond pulses are now available in a wide frequency range. In such experiments a probe is excited and the subsequent relaxation is studied. One example is the ultrafast dynamics of doubly occupied sites in the photo-excited quasi-2$d$ transition-metal dichalcogenide 1\emph{T}-TaS$_2$ \cite{1T-TaS2}.
Such excitations may even result in long-lived, metastable (``hidden'') states \cite{hidden}.

\section{Summary}

Dynamical mean-field theory (DMFT) is the generic mean-field theory for correlated electron systems and has shaped our current understanding of electronic correlations in solids \cite{Juelich-lecture-notes2}.
 In particular, the combination of DMFT with methods for the computation of electronic band structures provides a conceptually new framework for the realistic study of correlated materials. This approach and its various extensions make it possible to quantitatively understand and predict correlation phenomena in real materials ranging from complex anorganic systems all the way to biological matter.


\end{document}